# Coexisting Magnetic Order and Cooperative Paramagnetism in the Stuffed Pyrochlore $Tb_{2+x}Ti_{2-2x}Nb_xO_7$

B.G. Ueland[1], J.S. Gardner[1,2], A.J. Williams[3], M.L. Dahlberg[4], J.G. Kim[3], Y. Qiu[1,5], J.R.D. Copley[1], P. Schiffer[4], and R. J. Cava[3]

[1]*NIST Center for Neutron Research, National Institute of Standards and Technology, 100 Bureau Drive, Gaithersburg, MD 20899*

[2]*Indiana University, Bloomington, IN 47408*

[3]*Department of Chemistry and Princeton Materials Institute, Princeton University, Princeton, NJ 08540*

[4]*Department of Physics and Materials Research Institute, Pennsylvania State University, University Park PA 16802*

[5]*Department of Materials Science and Engineering, University of Maryland, College Park, MD 20742*

## Abstract

Neutron scattering and magnetization measurements have been performed on the stuffed pyrochlore system $Tb_{2+x}Ti_{2-2x}Nb_xO_7$. We find that despite the introduction of chemical disorder and increasingly antiferromagnetic interactions, a spin glass transition does not occur for $T \geq 1.5$ K and cooperative paramagnetic behavior exists for all $x$. For $x = 1$, $Tb_3NbO_7$, an antiferromagnetically ordered state coexisting with cooperative paramagnetic behavior is seen without applying any external fields or pressure, a situation advantageous for studying this cooperative behavior.

[*]*bgueland@nist.gov*

PACS numbers:  75.50.Mm, 75.50.Ee, 75.40.Cx, 75.25.+z

Spin liquids and their large spin classical analogues, cooperative paramagnets, posses strongly correlated spins (i.e. total magnetic moments) that cooperatively fluctuate down to zero temperature [1,2,3]. Recent studies of these paramagnets have focused on geometrically frustrated magnets with spins residing on corner sharing triangles [4,5,6,7], because such magnets remain paramagnetic down to temperatures much lower than those expected from the strengths of their spin-spin interactions due to both the frustration induced by the spatial arrangement of their spins [8,9,10] and the effects of reduced dimensionality of the magnetic sublattice. On the other hand, cooperative paramagnetic behavior has also been observed at temperatures below $T \approx 100$ K in the geometrically frustrated pyrochlore $Tb_2Ti_2O_7$, which has a 3-D magnetic sublattice of spins residing on corner sharing tetrahedra [11,12,13]. The cooperative paramagnetic behavior is evidenced by correlated spin fluctuations that persist below $T = 0.05$ K, and a lack of magnetic order at temperatures near the Weiss temperature of $\theta_W \approx -18$ K [11,14,15,16,17], which is indicative of the strength of the effective interaction between spins. Since the magnetic frustration arises from the arrangement of the spins, it may be expected that altering the lattice would drastically change the low temperature state. However, experiments studying the effects of magnetic dilution by substituting non-magnetic $Y^{3+}$ for Tb have shown that cooperative paramagnetic behavior persists in $Tb_{2-x}Y_xTi_2O_7$, but that the timescale of paramagnetic fluctuations depends on the amount of Y present [18]. Here, we take the opposite approach and examine the effects of increasing the number of spins in the lattice through magnetization and neutron scattering experiments studying the stuffed cooperative paramagnet $Tb_{2+x}Ti_{2-2x}Nb_xO_7$, in which $Tb^{3+}$ and $Nb^{5+}$ are stuffed into the lattice, randomly replacing $Ti^{4+}$ as $x$ is increased.

Similar to the stuffed spin ices $(Ho,Dy)_{2+x}Ti_{2-x}O_{7-x/2}$, $0 \leq x \leq 0.67$ [19,20,21], in $Tb_{2+x}Ti_{2-2x}Nb_xO_7$ as the stuffing, $x$, is increased, magnetic $Tb^{3+}$ is substituted for non-magnetic $Ti^{4+}$, placing the excess Tb at the vertices of an equivalent interpenetrating network of corner sharing tetrahedra, gradually transforming the magnetic sublattice from corner sharing to edge sharing tetrahedra, which is another frustrating geometry. In the materials studied here, non-magnetic $Nb^{5+}$ is substituted along with the stuffed Tb to maintain structural stability and O stoichiometry. The fully stuffed, $x = 1$, material, $Tb_3NbO_7$, with a chemically disordered fluorite lattice, has a magnetic sublattice of edge sharing tetrahedra randomly populated with spins, such that it contains 75 % Tb and 25 % Nb. We find that despite the introduction of chemical disorder and increasingly antiferromagnetic (AFM) effective interactions with increasing $x$, cooperative paramagnetic behavior similar to that in $Tb_2Ti_2O_7$ exists for all $x$ below $T = 10$ K, and that these materials do not freeze into a spin glass state for $T \geq 1.5$ K. For $x = 1$, $Tb_3NbO_7$, we unexpectedly find that long range AFM order coexists with cooperative paramagnetic behavior below $T_N = 2.2$ K.

Polycrystalline samples of $Tb_{2+x}Ti_{2-2x}Nb_xO_7$, $x = 0, 0.2, 0.4, 0.6, 1.0$, were synthesized using standard techniques [22] and confirmed to be single phase by X-ray diffraction. The magnetization $M$ was measured down to $T = 1.8$ K and in applied magnetic fields up to $\mu_0 H = 5.5$ T using a superconducting quantum interference device magnetometer. Inelastic neutron scattering measurements were made on $x = 0.2$ through 1.0 samples using the Disk Chopper Spectrometer, DCS, located at the Center for High Resolution Neutron Scattering, at the NIST Center for Neutron Research (NCNR) using neutrons with incident wavelengths of either $\lambda = 5$ Å or 1.8 Å. Neutron powder diffraction experiments were performed on $Tb_3NbO_7$ with the BT1 high resolution powder diffractometer at the NCNR utilizing the Ge (311) monochromator and 15′ in pile collimator. For the scattering experiments, the samples were placed in Al containers, and cooled in a $^4$He cryostat down to $T = 1.5$ K.

In Fig. 1a, we plot $\theta_W$ as a function of $x$ as obtained from Curie Weiss fits to $M(T)$ data taken in a field of $\mu_0H = 0.1$ T, examples of which are shown in Fig. 1b. (Uncertainties represent one standard deviation, either in the measured data or in fits to the data, and, unless indicated, are within the size of the symbols.) The determined effective moments are consistent with the value for free $Tb^{3+}$ for all $x$. Fig. 1a shows that $\theta_W$ monotonically decreases with increasing $x$, indicating that the effective magnetic interactions become increasingly AFM with stuffing, similar to the stuffed spin ices[19,21]. The $x = 0$ fit yields a value for $|\theta_W|$ smaller than that quoted above [11,15], however our fits were performed to lower temperature data 15 K < T < 25 K. Fitting to higher temperature data yields a value of $|\theta_W| \approx 18$ K for $x = 0$.

Figure 1c shows $M(\mu_0H)$ at $T = 2$ K for all $x$. While data for $x = 0$ continue to increase with field through $\mu_0H = 5.5$ T, likely due to the spacing of the crystal field levels [15], data for $x > 0$ have a shallower slope, trending towards saturation above $\mu_0H = 4$ T, and reaching almost half of the value for $Tb^{3+}$. In Fig. 1d, we plot the dc susceptibility, $\chi$, of $Tb_3NbO_7$ in a field of $\mu_0H = 0.01$ T. Data were taken after cooling in zero magnetic field (ZFC) or in the applied field (FC). Both ZFC and FC data show Curie type behavior down to $T < 10$ K, while, as shown in Fig. 1e, a peak occurs in the ZFC data at $T = 2.2$ K. While this difference between the ZFC and FC data may be characteristic of a spin glass transition [23], the neutron scattering experiments discussed below associate the peak with AFM long range order occurring at a Neel temperature of $T_N = 2.2$ K. No evidence for magnetic order or a spin glass transition is seen in our data for other $x$.

A cut along the energy transfer, $E$, of $I(Q,E)$, where $Q$ is the momentum transfer and $I$ is the intensity of the scattered neutron beam, from $\lambda = 5$ Å DCS data taken at $T = 50$ K, 10 K, and 1.5 K for $x = 0.2$ is shown in Fig. 2a after summing over 1 Å$^{-1} < Q < 1.2$ Å$^{-1}$. A sharp peak occurs at $E = 0$ meV due to elastic scattering, described below, and a temperature dependent peak near $E = 0.25$ meV occurs in the $T = 1.5$ K data. From $I(Q,E)$, we can identify this mode as $Q$ independent and associate it with the crystal field level seen in pure $Tb_2Ti_2O_7$ at $E \sim 1.8$ meV [14,15,16]. For $x = 0.2$, this mode has shifted to lower energy, presumably due to the change in the local environment of the Tb. In Fig. 2b we plot similar cuts of $I(Q,E)$ data along $E$ at $T = 1.5$ K for each $x$. The mode moves towards $E = 0$ meV with increasing $x$ up to $x = 0.4$ and disappears for higher $x$, indicating that changes to the local environment affect the crystal field splitting of the Tb levels. Figures 2c and 2d show cuts along $Q$ of $I(Q,E)$ data after summing over -0.1 meV $< E < 0.1$ meV. Figure 2c shows the temperature dependence of these data for $x = 0.2$, while Fig. 2d shows $T = 1.5$ K data for all $x$. In Fig. 2c, the (111) structural Bragg peak due to the pyrochlore lattice is seen near $Q = 1.05$ Å$^{-1}$, while temperature dependent diffuse scattering beneath the peak, arising from spin-spin correlations, grows in intensity with decreasing temperature. While Fig. 2d shows that diffuse scattering is present at $T = 1.5$ K for all $x$, the (111) pyrochlore peak decreases with increasing $x$ for $x = 0.2, 0.4$ and 0.6, consistent with the change from a pyrochlore to fluorite lattice, but reappears in the $T$

= 1.5 K data for $Tb_3NbO_7$. This allows us to identify it as a magnetic Bragg peak for $x = 1$.

In Fig. 3, we plot elastic neutron scattering data versus $Q$ due to spin-spin correlations. Data were initially obtained by summing $\lambda$ = 1.8 Å DCS data over -0.1 meV < $E$ < 0.1 meV. We then used $T$ = 50 K data as the nuclear background, subtracted them from the $T$ = 1.5 K data, and then divided by the square of the magnetic form factor for $Tb^{3+}$ yielding the scattering due to spin-spin correlations. Data for $x$ = 0.2, 0.4, 0.6, and 1 are shown in Fig. 3a, 3b, 3c, and 3d, respectively, and, as for $Tb_2Ti_2O_7$, were fit to a model describing isotropic nearest neighbor spin correlations, $I \sim \sin(R_0Q) / R_0Q$ [11]. Here, the spatial extent of the correlations is given by $R_0$, and the average value of $R_0(x)$ from the fits was determined to be $R_0$ = (3.76 ± 0.08) Å, which is close to the $x$ = 0 value[11], showing that stuffing does not affect the length of the spin-spin correlations.

Figure 4a shows BT1 powder diffraction data taken at $T$ = 50 K for $x$ = 1 along with results from Rietveld refinements performed using the FullProf software suite [24]. The refinements indicate that the sample has a fluorite crystal structure with space group Fm-3m and a lattice parameter of $a$ = 5.2837(2) Å. An impurity pyrochlore phase making up less than 1% of the sample was also fit to the data. $T$ = 1.5 K data containing magnetic Bragg peaks are plotted in Fig. 4b. From representational analysis performed using the program SARAh [25], along with magnetic Rietveld refinements, these data were found to be best described by the $\Gamma_5$ representation with propagation vector $\mathbf{k}$ = (½ ½ ½). The refinements yielded an ordered moment per spin of $|m|$ = 2.5(3) $\mu_B$, which is less than the value of $|m|$ = 9.6 $\mu_B$, expected for an isolated $Tb^{3+}$ and the reduced moment of $|m| \sim 5$ $\mu_B$ determined from calculations for $Tb_2Ti_2O_7$[15]. The inset to Fig. 4b contains a diagram of the magnetic order assuming each site is occupied by a spin, while the inset to Fig. 4a shows a diagram of the chemically disordered Tb/Nb sublattice. The values of chi squared for the $T$ = 50 K and 1.5 K data are $\chi^2$ = 1.72 and 2.27, respectively, indicating our model fits the data quite well.

From our experiments, we conclude that despite $\theta_W$ becoming increasingly AFM and the addition of chemical disorder with stuffing, $Tb_{2+x}Ti_{2-2x}Nb_xO_7$ does not freeze into a spin glass for $T \geq 1.5$ K. Furthermore, cooperative paramagnetic behavior in these materials is quite robust, with a similar correlation length for all $x$, suggesting that increasing the number of nearest neighbors, thus the effective AFM interactions, does not preclude the cooperative paramagnetic behavior. On the other hand, the fact that AFM order is seen for $x$ = 1 indicates that changes to the effective magnetic interaction are sufficient to induce magnetic order. We note that the change in energy of the first excited crystal field level with $x$ suggests that the proximity of the ground and excited state doublets in $Tb_2Ti_2O_7$ may not be crucial for the cooperative paramagnetic behavior. Also, the fluorite structure of $Tb_3NbO_7$ produces a cubic oxygen coordination around the $Tb^{3+}$ ions, which is expected to yield a non-magnetic singlet single ion ground state. Hence, any moment at the Tb site may be induced and not directly due to the angular momentum of the $Tb^{3+}$ ion.

Materials with spin liquid or cooperative paramagnetic behavior are also quite sensitive to the balance between lattice and magnetic energies. In particular, recent experiments on $Tb_2Ti_2O_7$ show that the development of lattice fluctuations coincides with the onset of cooperative paramagnetic behavior [26], and studies of $Tb_{2-x}Y_xTi_2O_7$ show that cooperative spin relaxation occurring at relatively high temperatures and high magnetic fields is dependent on the amount of dilution [27]. Additionally, experiments on $Tb_2Ti_2O_7$ show that applying a strong enough magnetic field at temperatures around $T = 3$ K induces long range magnetic order and condenses much of the diffuse scattering [28], while other experiments show that applying enough hydrostatic pressure creates a magnetically ordered state that coexists with liquid like fluctuations [29,30]. Similarly, in the spin liquid $Gd_3Ga_5O_{12}$, liquid like fluctuations are seen at low temperature while AFM order can be induced by a large enough magnetic field [4,6], and in $SrCr_{9x}Ga_{12-9x}O_{19}$ spin liquid behavior is found to coexist with spin glass behavior under an applied field [5,7]. Here, we show coexisting magnetic order and liquid type behavior in $Tb_3NbO_7$ without applying any pressure or magnetic fields, a situation conducive to performing much simpler experiments. In addition, the change in crystal structure with stuffing should facilitate future studies into what role the lattice plays in magnets with cooperative paramagnetic behavior.

One possible explanation for the coexistence of both magnetic order and cooperative paramagnetic behavior is that spatially separate regions of AFM static order and strongly correlated fluctuating spins exist. For example, in the stuffed spin ices, it has been suggested that the coexistence of finite entropy and ac magnetic susceptibility as $T \sim 0$ K may be due to pyrochlore structural domains [21,31] or separated unfrozen regions of spins [20]. However, then the magnetic Bragg peaks in Fig. 4b would not be resolution limited, but rather spread out in $Q$ (i.e. $2\theta$). Future experimental and theoretical studies examining the size of the fluctuating moment, the spin lattice coupling, the effects of quantum fluctuations [32], and the timescale of spin fluctuations in $Tb_{2+x}Ti_{2-2x}Nb_xO_7$, should shed light on the conditions necessary for persistent cooperative paramagnetic behavior and, for $x = 1$, coexistent long range order. More generally, future studies on $Tb_{2+x}Ti_{2-2x}Nb_xO_7$, and similar materials will enhance the understanding of competing energies in highly correlated materials.

**ACKNOWLEDGEMENTS**

We gratefully acknowledge helpful discussions and exchanges with M. A. Green, P. Zajdel, and J. W. Lynn, support from the NRC/NIST Postdoctoral Associateship Program, and NSF grants DMR-0454672, DMR-0701582, and DMR-0703095.

Figure captions

FIG. 1. (a) $\theta_W(x)$ determined from Curie-Weiss fits to $\chi^{-1}(T)$. The line is a guide to the eye. (b) $\chi^{-1}(T)$ for $x = 0.2$ and 1 and the corresponding Curie-Weiss fits. (c) $M(\mu_0 H)$ for each $x$ at $T = 2$ K. (d) Zero field cooled (ZFC) and field cooled (FC) $\chi(T)$ data for $x = 1$ at $\mu_0 H = 0.01$ T. (e) A blowup of the low temperature portion of the $x = 1$ $\chi(T)$ data.

FIG. 2. (a) $I(E)$ for $x = 0.2$ at $T = 50$ K, 10 K, and 1.5 K after summing between 1 Å$^{-1}$ < $Q$ < 1.2 Å$^{-1}$. (b) $I(E)$ as in (a) at $T = 1.5$ K for each $x$. Arrows indicate the approximate positions of crystal field exictations. (c) $I(Q)$ for $x = 0.2$ after summing between -0.1 meV < $E$ < 0.1 meV at $T = 50$ K, 10 K, and 1.5 K. (d) $I(Q)$ as in (c) at $T = 1.5$ K for each $x$.

FIG. 3. The scattered neutron intensity due to spin-spin correlations as a function of $Q$ for $x = $ 0.2 (a), 0.4 (b), 0.6 (c), and 1 (d), as described in the text. Lines are fits as described in the text. The rise at low $Q$ for all $x$ may be due to air scattering.

FIG. 4. $T = 50$ K (a) and $T = 1.5$ K (b) powder diffraction data for Tb$_3$NbO$_7$ and their refinements  The diagram in (a) shows the Tb/Nb sublattice of side sharing tetrahedra, while the diagram in (b) shows the AFM order. Top and bottom sets of tic marks indicate positions of the fluorite and pyrochlore Bragg peaks, respectively. The middle set of tic marks in (b) indicates the positions of the magnetic Bragg peaks. Uncertainties are statistical in nature and within the size of the symbols.

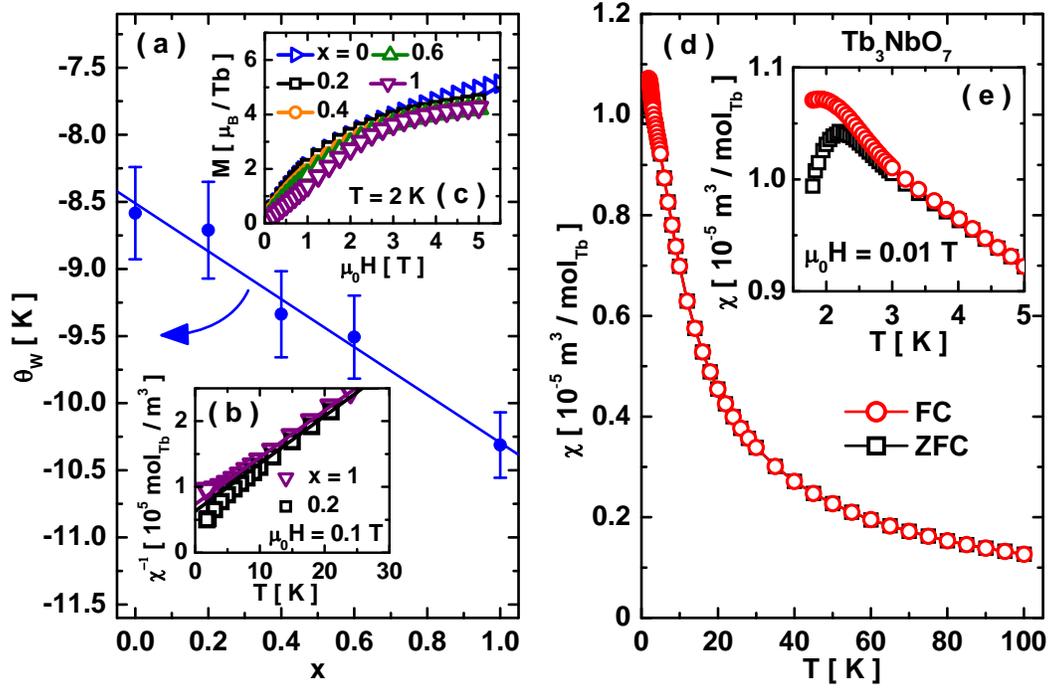

Fig. 2. Ueland *et al*.

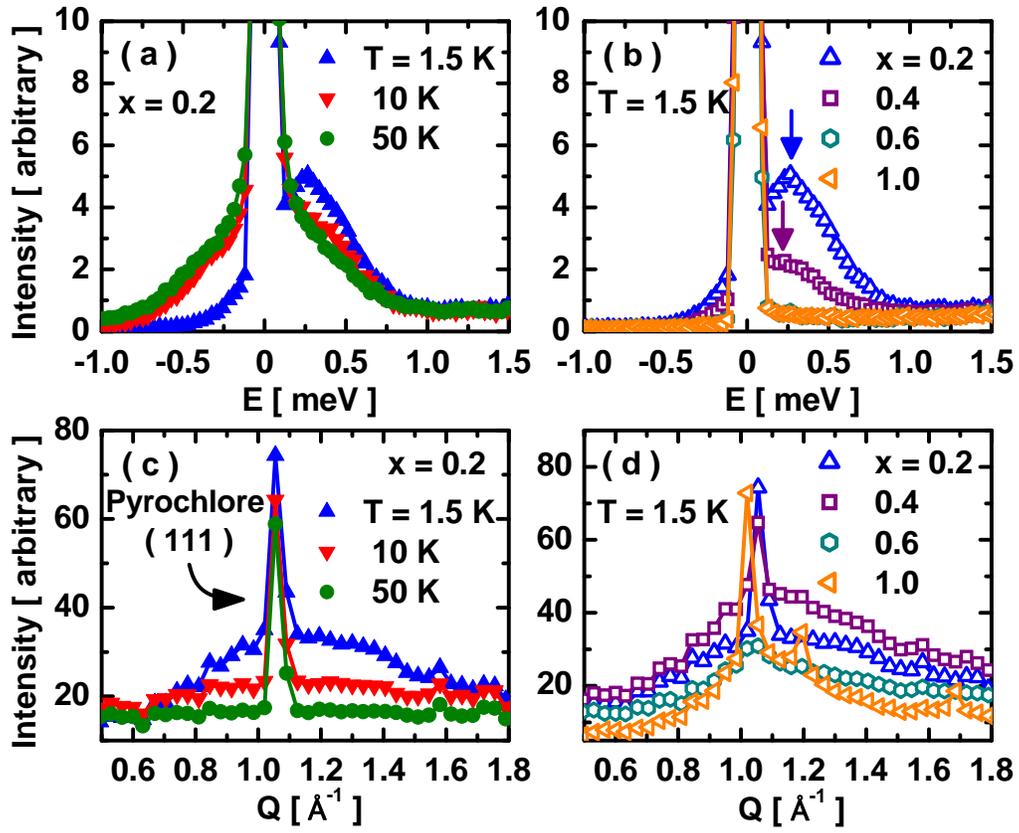

Fig. 3. Ueland *et al*.

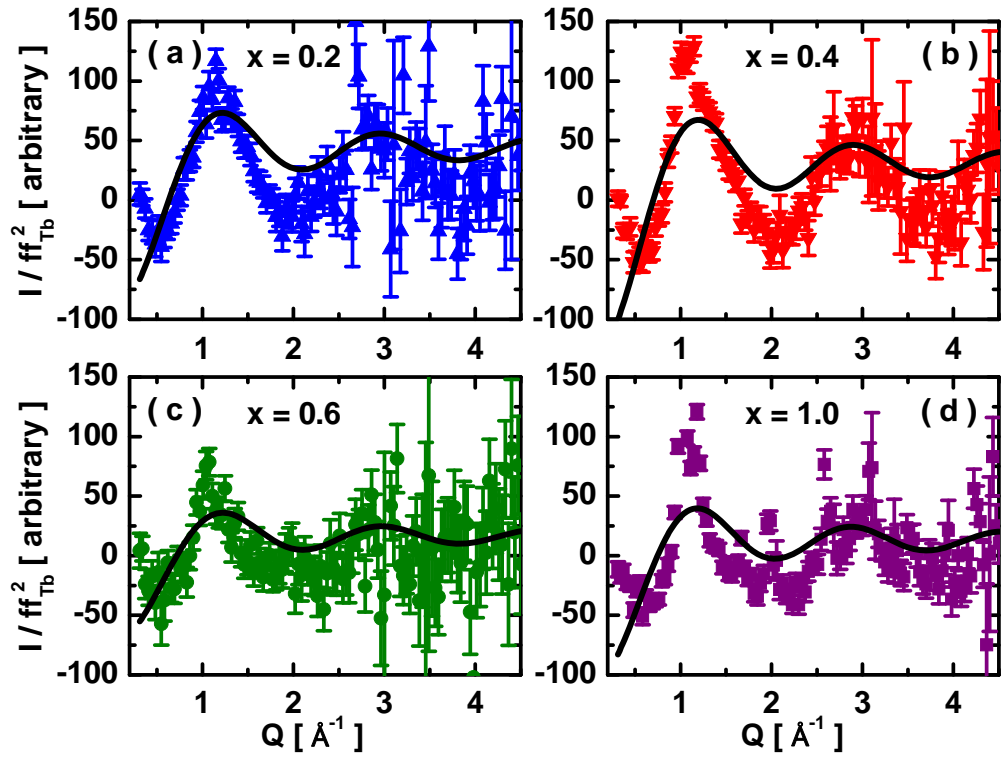

Fig. 4. Ueland *et al*.

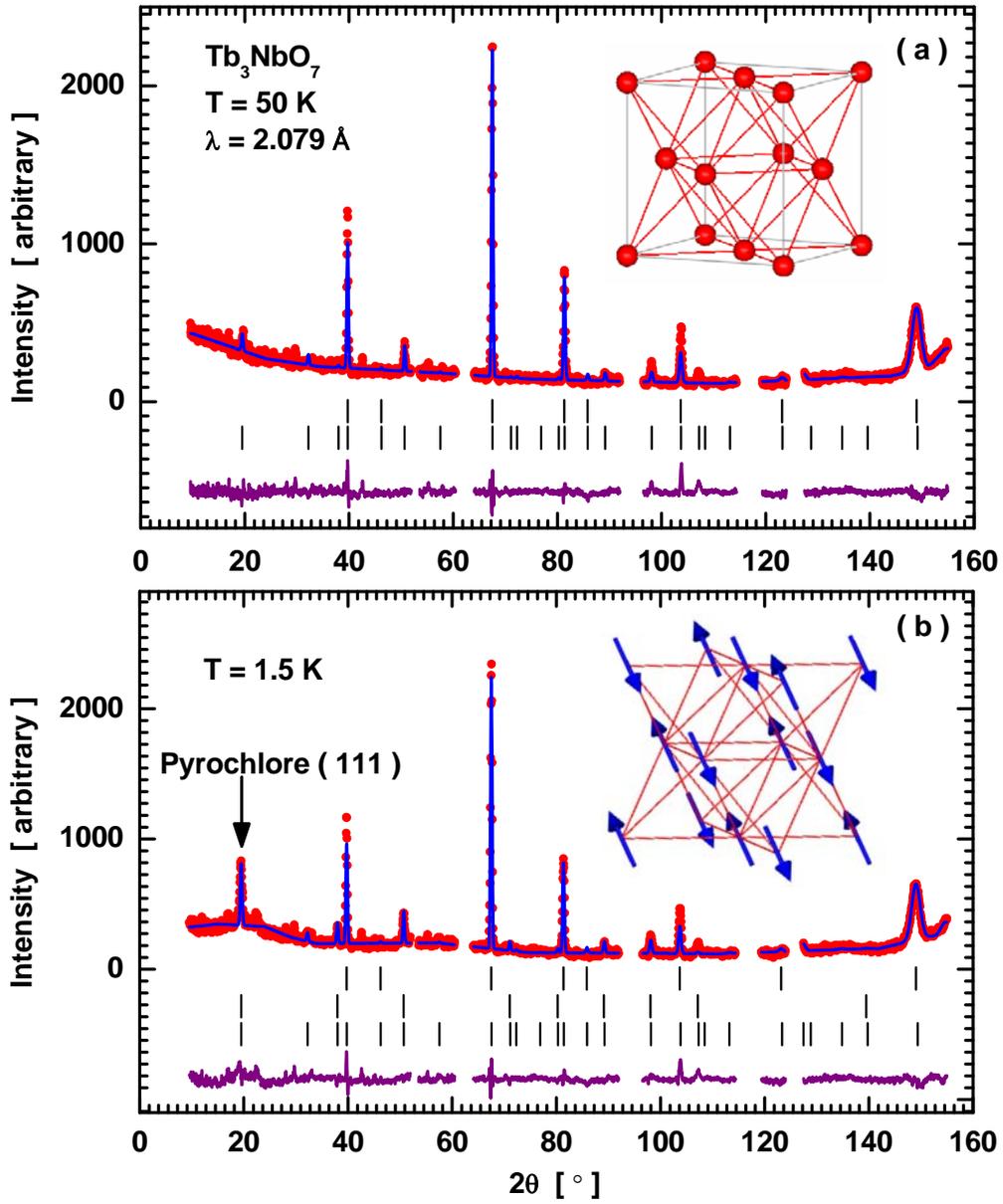